\documentclass{PoS}
\usepackage{cite}

\title{The Interaction Of AGN Jets With Obstacles}

\ShortTitle{AGN Jets and Obstacles}

\author{\speaker{N\'uria Torres-Alb\`a}\\
        Departament de F\'isica Qu\`antica i Astrof\'isica, Institut de Ci\`encies del Cosmos (ICC), Universitat de Barcelona (IEEC-UB), Mart\'i i Franqu\`es 1, 08028 Barcelona, Spain\\
        E-mail: \email{ntorres@fqa.ub.edu}}

\abstract{Extragalactic jets are launched from the innermost regions of galaxies, near the central supermassive black hole. As they propagate, they must cross the whole galaxy, and in this process they interact with a variety of obstacles; including gas clouds, populations of stars or even supernova remnants. The interaction between jets and penetrating obstacles has been studied as a possible method for jet mass-loading and deceleration, as well as of production of gamma-ray emission, through non-thermal particles accelerated in shocks. Interaction with individual objects, such as stars or gas clouds, can explain both rapid variability in blazars, and gamma-ray flares. Interaction with whole populations of obstacles, however, may lead to the production of persistent gamma-ray emission.}

\FullConference{High Energy Phenomena in Relativistic Outflows VII - HEPRO VII\\
		9-12 July 2019\\
		Facultat de F\'isica, Universitat de Barcelona, Spain}

\begin{document}

\newcommand{\apj}{ApJ}
\newcommand{\apjl}{ApJ Lett.}
\newcommand{\ijmpd}{Int. J. Mod. Phys. D}
\newcommand{\mnras}{MNRAS} 
\newcommand{\aap}{A\&A}
\newcommand{\apjs}{ApJS}
\newcommand{\nat}{Nature}
\newcommand{\nar}{New Astron. Rev.}
\newcommand{\araa}{ARA\&A}
\newcommand{\apss}{Ap\&SS}
\newcommand{\aaps}{A\&AS}
\newcommand{\pasj}{PASJ}
\newcommand{\physrep}{Physics Reports}
\newcommand{\ssr}{Space Sci. Rev.}
\newcommand{\sovast}{Soviet Ast.}
\newcommand{\aapr}{A\&A Rev.}

\section{Introduction}
 
Supermassive black holes, present in the innermost regions of galaxies, may accrete the material surrounding them, becoming active galactic nuclei (AGN). Some AGN produce collimated relativistic outflows, or jets \cite{Begelman1984},
which propagate through the host galaxy. This propagation will inevitably lead to the jet interacting with a variety of obstacles,  which may include stars, gas, and dense clouds. Through this interaction, both jet and obstacles can be dynamically affected, and non-thermal radiation might be produced. 

Numerous works have studied these interactions, considering obstacles in the innermost regions of the galaxy, such as broad-line region (BLR) \cite{Dar1997,Beall1999,Araudo2010,delPalacio2019} and narrow-line region clouds (NLR) \cite{Steffen1997}. At $\sim$pc and $\sim$kpc scales, either individual stars with high mass-loss rates \cite{Barkov2010,Araudo2013,Khangulyan2013} or whole populations of stars \cite{Bednarek1997, Bosch-Ramon2015,Vieyro2017,Perucho2014,Torres-Alba2019} and supernova remnants \cite{Vieyro2019} have been considered. At larger scales, when the jet reaches the halo, it might interact with globular clusters \cite{Bednarek2015}, interstellar medium (ISM) clouds \cite{Jeyakumar2005,Choi2007})and even intergalactic medium (IGM) clouds \cite{Higgins1999}.

Evidences  from  jet  interaction  with  the  medium  mainly arise  from two observational facts: the Fanaroff-Riley type I/II (FRI/II) dichotomy \cite{Fanaroff1974}, and the presence of knots, or localised intensity enhancements, within the jet structure (e.g. in M87 \cite{Wilson2002} or Centaurus A \cite{Hardcastle2007}). For a recent and thorough review on the topic of the interaction between AGN jets and their ambient medium see \cite{Perucho2019}.

\section{Dynamical impact of obstacles on AGN jets}

Based on their large-scale morphology in radio, Fanaroff and Riley \cite{Fanaroff1974} separated AGN jets into two categories. FRII jets propagate up to hundreds of kiloparsec and are brighter at the edges, point at which the density of the IGM is sufficient to stop their propagation. The collision with the IGM results in the formation of shocks, which are prominent non-thermal emitters. Up to these regions, or hotspots, FRII jets remain collimated and highly relativistic. On the other hand, FRI sources show plumed or irregular morphologies, and are typically decelerated to non-relativistic speeds within the first few kpc of the host galaxy. As a result of this deceleration, they are brighter closer to the nucleus of the galaxy, and present decollimation.

The FRI/II dichotomy has been related to a difference in jet power, with FRIIs typically having $L_{\rm j}\geq 10^{45}$ erg s$^{-1}$ and FRIs corresponding to $L_{\rm j}\leq 10^{44}$ erg s$^{-1}$ \cite{Ghisellini2001}. It is thought that interaction with the ISM plays a role in determining jet morphology, as the weaker FRI jets can be decelerated due to entrainment of external material.  As  the  jet  propagates,  the  difference  of velocity between its bulk motion and the surrounding medium gives raise  to  a  shear  layer,  in  which  small  instabilities  can  develop  and grow, resulting in significant mixing. The exchange of momentum in this process forces the jet to decelerate significantly, and can even disrupt it \cite{deYoung1993,Bicknell1994,Wang2009}. Mass entrainment can also occur through the presence of standing recollimation shocks. As the jet expands, its pressure decreases until, at  some  point,  it  becomes  lower  than  that  of  the  ambient  medium. The jet is forced to recollimate when colliding with the ISM, which can  produce  inner  shocks  that  result  in  deceleration  and  facilitate mixing of ambient material \cite{Perucho2007}. A similar effect is produced when the jet encounters density variations in the ISM \cite{Meliani2008}.

\subsection{Jet disruption and deceleration}

\begin{figure}
\center
\includegraphics[width=\textwidth]{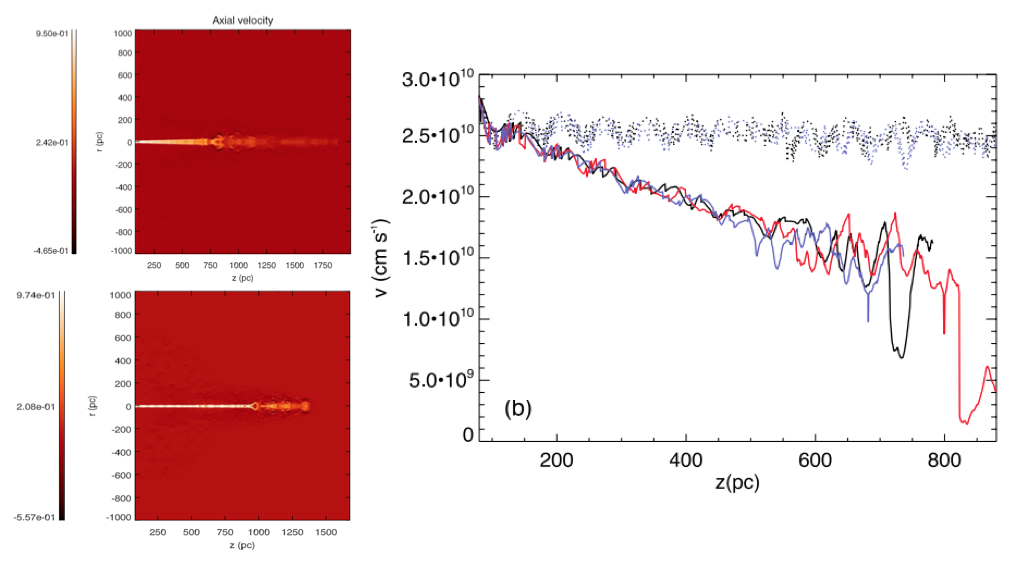}
\caption{\textit{Right}:  Axial velocity plots of identical jets with (top) and without (bottom) mass-loading. \textit{Left}: Average value of the axial velocity as a function of distance for jets with $L_{\rm j}=5 \times 10^{41}$ erg s$^{-1}$ and no/little mass loading (dotted lines), compared to jets with the same power and realistic values of mass loading (solid lines). Figure courtesy of Manel Perucho \cite{Perucho2014}.}
\label{fig1}
\end{figure}

Another possibility for entrainment of external material lays in the mass ejected by stellar winds. Komissarov \cite{Komissarov1994} showed, through numerical estimations, that stellar populations in elliptical galaxies can decelerate the jets of weak FRI sources, a result supported by 2D simulations \cite{Bowman1996}. Perucho et al. \cite{Perucho2014} also showed through simulations that a realistic distribution of mass injected into a jet could decelerate weak jets ($L_{\rm j}\sim10^{42}$ erg s$^{-1}$) and reproduce the broadening of the opening angle observed in FRI sources (see Fig. \ref{fig1}). However, they found stellar winds to have no significant impact on jets with $L_{\rm j}\sim10^{44}$ erg s$^{-1}$. These results are consistent with the analysis performed by Laing \& Bridle \cite{Laing2014} using high-resolution imaging and polarimetry data from the Very Large Array (VLA). Modelling the radio emission, they found they could reproduce both intensity maps and velocity gradients of weak jets with the assumed stellar populations of their host galaxies (although they could not rule out ISM entrainment as the cause). However, powerful jets were found to be slower at the edges than on-axis, which implies the dominant deceleration mechanism is necessarily entrainment of ISM through the boundary layer of the jet. 

On the other hand, Vieyro et al. \cite{Vieyro2017} modelled the population of red giants present within the jet of M87 ($\sim10^6$ red giants at all times), and found that the mass injected by the whole population was likely enough to dynamically affect the jet ($L_{\rm j}\sim10^{44}$ erg s$^{-1}$) on scales of $\sim$3~kpc. Hubbard \& Blackmann \cite{Hubbard2006} also reached the conclusion that a small number of high-mass loss stars is likely to be more efficient at decelerating AGN jets than the rest of the stellar population. 

Strong asymmetries in jet morphology, such as a bending jet, have also been interpreted as a result of the interaction with ambient medium; in particular a collision between the jet and a dense ISM cloud (e.g., 3C 43 \cite{Cotton2003}). These asymmetries, both in morphology and brightness, have been reproduced in 2D and 3D simulations \cite{Jeyakumar2005,Choi2007} considering jets propagating through a medium filled with dense clouds.

\subsection{Acceleration of gas clouds by AGN jets}

\begin{figure}
\center
\includegraphics[width=0.9\textwidth]{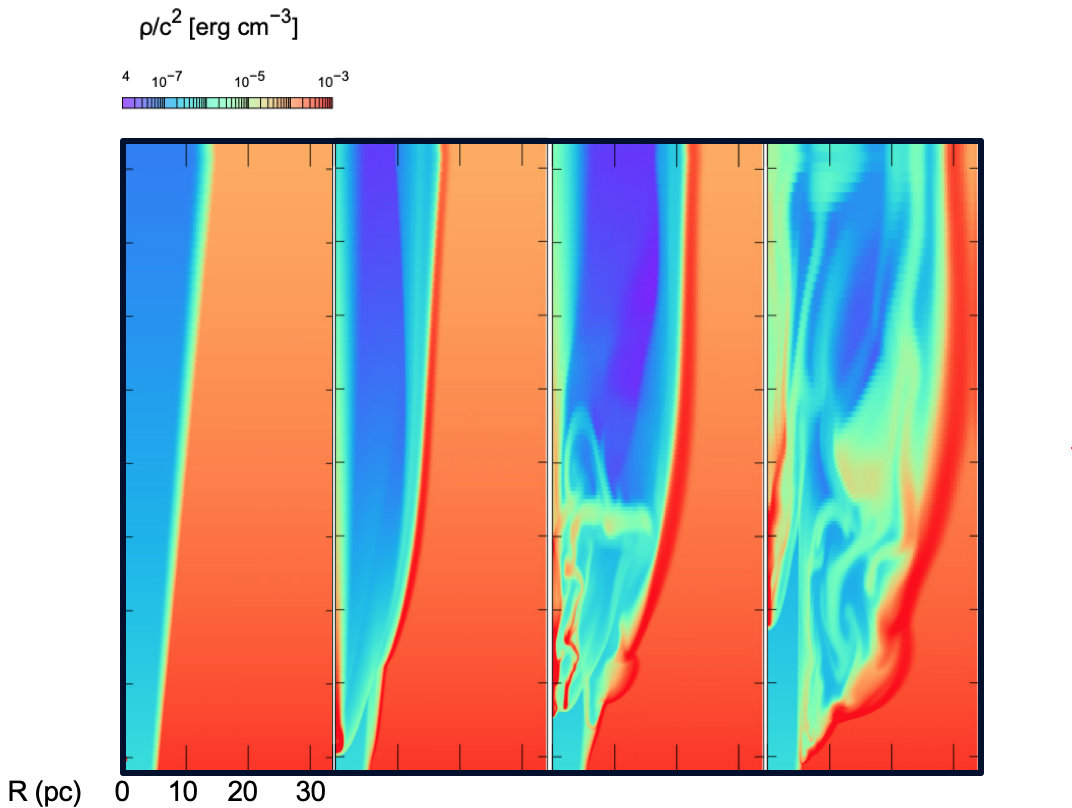}
\caption{Evolution of a SN ejecta expelled within the jet of an AGN, and its impact on the shear layer. Figure adapted from \cite{Vieyro2019}.}
\label{fig2}
\end{figure}

AGN jets also interact with gas clouds, from either the BLR or NLR, or even those introduced by stars at the moment they penetrate the jet \cite{Torres-Alba2019}. Any cloud entering the jet is impacted by its ram pressure, which results in the propagation of a shock through the cloud layers. The cloud can enter the jet if its orbital velocity is faster than the shock speed, else it is carried upwards through the shear layer. If it penetrates, the impact of the jet causes a transfer of momentum, which results in the acceleration of the cloud. In addition, the shock wave propagates through the cloud, compressing it and heating it up, which results in a lateral expansion (see e.g. \cite{Barkov2010,Perucho2017}). BLR/NLR clouds and the like are generally not massive enough to, in turn, also have a dynamical impact on the jet. However, if a supernova (SN) explosion takes place within it, up to 10~M$_\odot$ of material can be injected into it at once. As shown by recent simulations \cite{Vieyro2019}, such a large blob is also accelerated donwstream, but in this process its final radius becomes as large as the jet radius, which results in an impact against the shear layer (See Fig. \ref{fig2}). 

\section{Non-thermal emission}

Interaction with the environment not only leads to deceleration and/or disruption of AGN jets, but also to the production of non-thermal radiation. As an obstacle manages to penetrate the jet (i.e. a gas blob, a star), it is impacted by its ram pressure and a double bow shock (in which both jet and cloud/star material are shocked) is formed. 

In a shock, particles can be accelerated up to relativistic energies through various mechanisms, although first-order  Fermi  acceleration  (also  known  as  diffusive  shock  acceleration, DSA) is most commonly invoked as the dominant mechanism at play.  According to DSA, particles are continuously reflected by magnetic inhomogeneities, which scatter them upstream and downstream the flow, gaining energy with each crossing of the shock front, until they reach highly relativistic velocities (e.g. \cite{Bell1978,Blandford1987}). Particles can then either escape the flow (e.g. cosmic rays), or radiate their energy through non-thermal processes. 

Particle  acceleration  in  shocks  with  obstacles has  since  long  been proved capable of reproducing observational features, such as the presence of knots in the jets of AGN (e.g. \cite{Blandford1979}). In fact, radio observations of the inner pc of Centaurus A show morphologies reminding of a bow-shock shape, which can be reproduced assuming the presence of a high-mass-loss-rate star within the jet \cite{Muller2014}.

\subsection{TeV variability}

Dar \& Laor reproduced a very short and intense TeV gamma-ray flare of Mrk 421 (blazar) by modelling the emission produced in collisions between high-energy jet protons and BLR clouds crossing the line of sight close to the black hole \cite{Dar1997}, a process through which high-energy neutrinos can also be generated, and both delayed optical and X-ray flares produced. Similarly, Araudo et al. \cite{Araudo2010} explained both fast TeV variability and steady emission in non-blazar AGN, a result later extended to blazar sources \cite{delPalacio2019}. Interaction between the jets and NLR clouds, on the other hand, has been shown to affect the emission-line brightness and velocity distribution of the clouds \cite{Steffen1997}, which is also in accordance with observations of Seyfert galaxies such as NGC 1068 \cite{Capetti1997}.

Rapid TeV variability can also be explained assuming jet-star interactions, as long as the star is sufficiently evolved (i.e. red giant, AGB phase star) and it crosses the jet close enough to the base \cite{Barkov2010}. Analytical estimations show that it is possible to reproduce both light-curves and spectral energy distributions (SEDs) of flares (e.g. for M87 \cite{Barkov2012}, 3C 454.3 \cite{Khangulyan2013} or PKS 1222+21 \cite{Banasinski2016}). The latter assumes that a red giant crossing the jet can interact with it through its winds, causing a "plateau" state while it crosses (in timescales of $\sim$ weeks); while bright and short flares can be produced by a few clouds of matter lost by the red giant after the initial impact of the jet. These clouds would expand and be quickly accelerated downstreams, emitting more intensely in the process. 

\subsection{Persistent emission}

One expects, however, that the largest number of interactions with stars will take place at much larger heights, where the jet has significantly broadened. Then, the whole population of stars can produce a series of bow-shocks within the jet, which added together results in significant emission. Vieyro et al. \cite{Vieyro2017} modelled the high-mass-loss-rate stellar populations present in both intensely star-forming galaxies (luminous infrared galaxies, or LIRGs, with infrared luminosities $L_{\rm IR}\geq10^{11}$ L$_\odot$) and in elliptical galaxies. In the former, OB stars are expected to form in molecular disks within the inner few hundreds of parsecs of the galaxy, while in the latter millions of bulge red giants can be present within the jet at all times. As a result of this interaction, persistent gamma-ray emission can be detectable in both $z\sim0.1$ blazars and nearby ellipticals.

Whole populations of stars have also been found to be able to reproduce the X-ray emission of the jet of Centaurus A. At a distance of $\sim$10~Mpc, the jet of Centaurus A is one the most well-studied, with the \textit{Chandra} data identifying numerous individual knots within the central kpc of the galaxy. Wykes et al. \cite{WykHar2015} modelled the winds and distribution of the known stellar populations present in the galaxy, accounting for both red giants and younger stars being formed in the dust lane that (presumably) crosses the jet. As a result, they were able to explain both the knotted and diffuse emission generated in X-rays solely through the stellar populations present in the galaxy.

Persistent emission can also be produced by populations of stars penetrating the jet \cite{Torres-Alba2019}. As stars propagate through the ISM, their own orbital speeds cause them to impact the ambient gas, which accumulates around them in a "bubble" of material. This bubble is carried along with the star until it reaches the jet, moment at which it is impacted by its ram pressure. As mentioned above, if the star is fast enough, part of this bubble manages to enter the jet and is expelled as a roughly spherical blob inside. If enough stars penetrate the jet per unit of time, the multiple clouds being accelerated downstream at different jet heights can result in significant persistent gamma-ray emission. This, however, is likely to be the case only for red giants in elliptical galaxies, and through leptonic synchrotron emission (which implies that particles first need to be accelerated to high enough energies). 

\subsection{Longer-lasting transient events}

A massive red giant, or an AGB star entering the jet, could however introduce a bubble massive enough for it to dominate over the emission of the whole population (caused by both the introduction of bubbles, and the population of stars always present within the jet, interacting with it through their winds). In such a case the interaction would result in a brighter phase, lasting from a few months to tens of years, resulting in long-term variability.

\begin{figure}
\center
\includegraphics[width=0.9\textwidth]{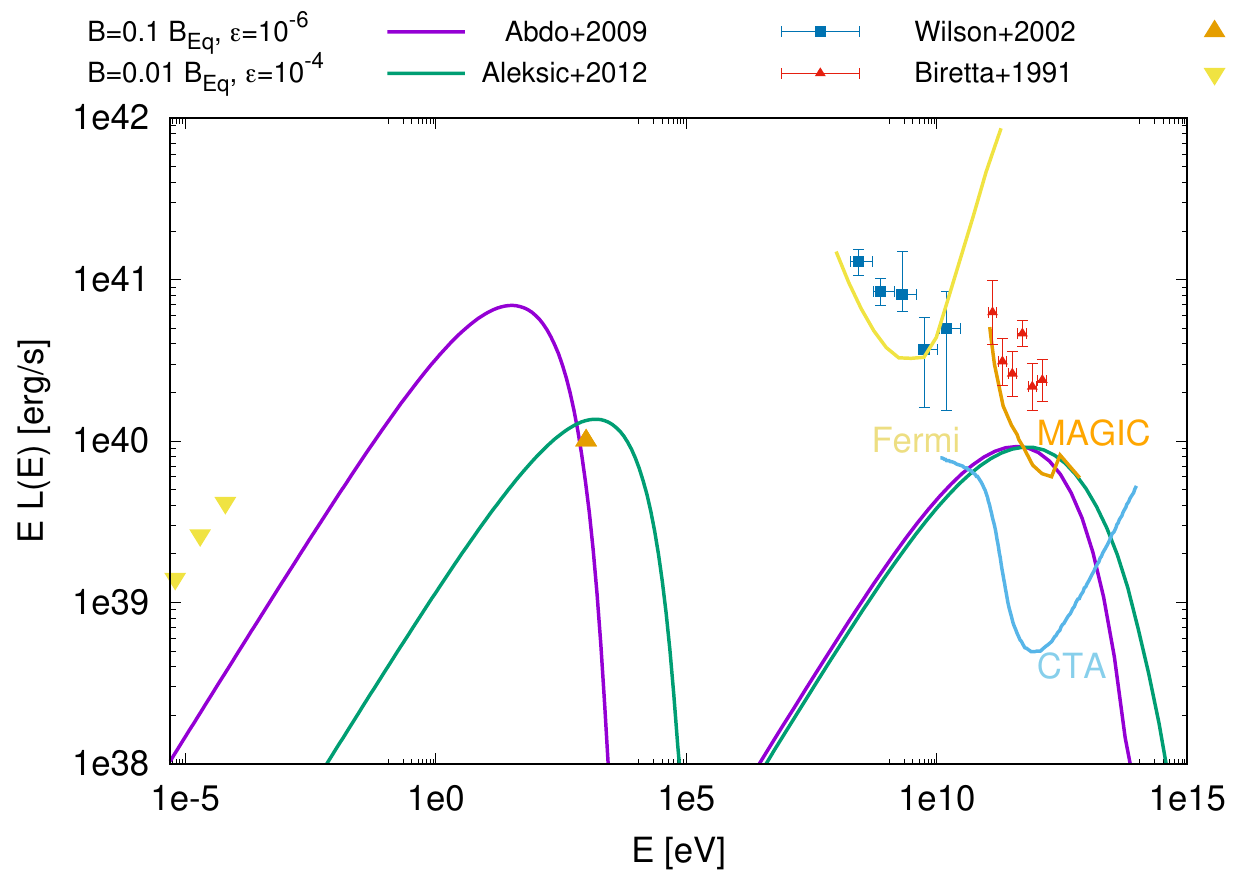}
\caption{Synchrotron and IC SEDs for the interaction between a type Ia SN remnant and the jet of M87, for different values of the magnetic field, $B$ (as a function of the equipartition value), and the particle acceleration efficiency, $\epsilon$. Also plotted are sensitivity curves for Fermi \cite{Fermi2012}, MAGIC \cite{MAGIC2016} and CTA (both after 50h of direct observation), as well as MAGIC data from \cite{magic2012M87}, {\it Fermi} data from \cite{abdo2009M87}, X-ray data from \cite{wilson2002M87} and radio data from \cite{DoeFis2012}.  X-ray and radio data both correspond to emission at kpc scales, taken from the brightest knot (knot A) in the jet of M87, and have error bars smaller than the size of the point. The resolution of gamma ray observatories does not allow to disentangle emission from the kpc-scale jet and the nucleus. All data were taken during the source's low-emission state.}
\label{fig3}
\end{figure}

A supernova (SN) explosion within the jet would also result in a significant enhancement of non-thermal emission, much above that expected from the interaction with the stellar population. Results by Vieyro et al. \cite{Vieyro2019}, who studied core-collapse SNe in star-forming galaxies, show that this emission would be detectable for nearby radio galaxies, and for powerful blazars up to $z\sim1$. In the case of elliptical galaxies, in which Type Ia SNe are frequent, results show that up to $\sim100$ persistent VHE sources could be detected in the whole sky, once CTA is operative (Torres-Albà \& Bosch-Ramon, in prep. See Fig. \ref{fig3}). In both cases, the emission would last thousands of years.

\section{Distinguishing between different obstacles as the source of non-thermal emission}

Interaction between AGN jets and various obstacles/ISM, as well as scenarios such as the spine sheath \cite{Tavecchio2016} or jet-in-jet \cite{Giannios2009} models, can explain production of non-thermal gamma-ray emission in AGN jets. However, distinguishing between any of these processes as the origin of the emission in a particular source is non-trivial. 

For nearby sources, X-ray and radio imaging resolution might be enough to disentangle morphological signatures of the emission. Unlike stellar populations or wind bubbles, which would result in multiple, small knots of different brightness, a SN ejecta is expected to cover the whole jet section for the majority of its interaction with the jet. This allows  to rule out single bright knots with a too-small section in nearby radio-galaxies. Furthermore, both wind bubbles and SNRs are expected to move downstream of the jet, while a star (or group of stars) within the jet would cross it roughly horizontally. Both the crossing time and the SNR evolution are expected to last thousands of years, meaning they would appear stationary in observations. A  wind  bubble  (say,  of $\sim10^{-6}$ M$_\odot$, easily produced by a red giant of 1 M$_\odot$ penetrating the jet)  would be  accelerated  quickly  enough  to  appreciate,  in  timescales  of  years, its movement.

Distinction between stationary and moving knots has been applied to Centaurus A using \textit{Chandra} data spanning 15 yr \cite{Snios2019}. As a result, Snios et al. can associate unmoving knots to crossing stars (i.e. AGB stars), and suggest the moving knots could be the result between the collision of the jet plasma and an obstacle on the verge of being dissolved in the jet (i.e. molecular clouds). Estimations of the magnetic field \cite{Goodger2010} allow to rule out magnetic reconnection (also known as impulsive particle acceleration) as the origin of the stationary knots, as it would imply a drastic change of appearance in the knots in about six years, which is not observed.

Sources like Centaurus A allow for observations with high sensitivity and high resolution at different wavelengths, as well as for very accurate modelling of both the jet (e.g. magnetic field) and host galaxy (e.g. stellar ages and distribution) properties. However, the large majority of VHE sources are at distances that make such distinctions impossible. 

For sources at a known distance, and with a minimum knowledge of host galaxy properties (e.g. galaxy type, mass) it is possible to rule out some of the mentioned mechanisms as the origin of the emission. For example, for a given estimation of the basic characteristics of a stellar population, the emission generated by a SNR being accelerated by the jet will always be larger than the emission generated by the parent population. In a such a way, it is possible to estimate which is the most likely source of the emission between these two options. Monitoring the variability would then allow to distinguish between bubbles dominating the emission (as long-term variability would eventually be involved), or the main source being the stars losing mass through their winds within the jet.

However, there are still many unknowns in the modelling of non-thermal emission of highly relativistic particles. It is not fully known yet if jets have an important hadronic component, and whether it can dominate the emission over that of the leptonic particles. The strength of the magnetic field with respect to the equipartition value (i.e. total magnetic energy density is equal to half the jet energy density), as well as the efficiency of particle acceleration are not well constrained, which results in a degeneracy when estimating synchrotron emission. Similarly, it is not known which fraction of jet energy goes into the acceleration of non-thermal particles. 

All of these unknowns make it likely that observations can be reproduced by more than one model, particularly for sources with unconstrained properties.

\section{Conclusions}

AGN jets are likely to interact with a multitude of obstacles as they propagate through the host galaxy, as shown both by analytical estimations and by observations. These objects, which include BLR and NLR clouds, stars, SNRs and ISM/IGM clouds can mostly result in both jet deceleration and disruption and the production of non-thermal gamma-ray emission. 

According to both simulations and modelling of observations, it seems likely that stellar populations (in particular red giants) can decelerate jets of up to $L_{\rm j} \simeq 10^{44}$ erg s$^{-1}$. More powerful jets must be decelerated through  other means, such as recollimation shocks or entrainment of ISM material through a turbulent shear layer.

Interaction with obstacles can reproduce both fast TeV variability (interaction with stars/clouds crossing the jet close to its base), long-term transient emission (stars penetrating the jet), emission enhancements lasting times of the order of thousands of years (SNe exploding within the jet) and persistent emission (winds of stellar populations impacting the jet, stars penetrating the jet, BLR clouds). Distinguishing between the different means of emission remains difficult, unless studying nearby sources (e.g. Centaurus A).

\bibliographystyle{JHEP}
\bibliography{Referencies}

\end{document}